\documentclass[showpacs,preprintnumbers,amsmath,pre,amssymb,preprint]{revtex4}

\usepackage{graphicx}% Include figure files
\usepackage[latin1]{inputenc}
\usepackage{dcolumn}% Align table columns on decimal point
\usepackage{bm}% bold math
\usepackage{epsfig}

\begin{document}

\title{Horizontal Visibility graphs generated by type-I intermittency}% Force line breaks with \\

\author{\'Angel M. N\'u\~{n}ez}
\affiliation{Dept. Matem\'{a}tica Aplicada y Estad\'{i}stica. ETSI Aeron\'{a}uticos, Universidad Polit\'{e}cnica de Madrid, Spain.}
\author{Bartolo Luque}
\affiliation{Dept. Matem\'{a}tica Aplicada y Estad\'{i}stica. ETSI Aeron\'{a}uticos, Universidad Polit\'{e}cnica de Madrid, Spain.}
\author{Lucas Lacasa}
\affiliation{Dept. Matem\'{a}tica Aplicada y Estad\'{i}stica. ETSI Aeron\'{a}uticos, Universidad Polit\'{e}cnica de Madrid, Spain.}
\author{Jose Patricio G\'{o}mez}
\affiliation{Dept. Matem\'{a}tica Aplicada y Estad\'{i}stica. ETSI Aeron\'{a}uticos, Universidad Polit\'{e}cnica de Madrid, Spain.}
\author{Alberto Robledo}
\affiliation{Instituto de F\'{\i}sica y Centro de Ciencias de la Complejidad, Universidad Nacional Aut\'{o}noma de M\'{e}xico, Mexico.}

\date{\today}% It is always \today, today,

%{Low-dimensional chaos}

\begin{abstract}
The type-I intermittency route to (or out of) chaos is investigated within the Horizontal Visibility graph theory. For that purpose, we address the trajectories
generated by unimodal maps close to an inverse tangent bifurcation and construct, according to the Horizontal Visibility algorithm, their associated graphs. We show
how the alternation of laminar episodes and chaotic bursts
has a fingerprint in the resulting graph structure. Accordingly, we derive a phenomenological theory that predicts quantitative values of several network parameters. In particular, we predict that the characteristic power law scaling of the mean length of laminar trend sizes is fully inherited in the variance of the graph degree distribution, in good agreement with the numerics. We also report numerical evidence on how the characteristic power-law scaling of the Lyapunov exponent as a function of the distance to the tangent bifurcation is inherited
in the graph by an analogous scaling of the block entropy over the degree distribution. Furthermore, we are
able to recast the full set of HV graphs generated by intermittent dynamics into a renormalization group framework, where the fixed points of its graph-theoretical RG flow
account for the different types of dynamics. We also establish that the nontrivial fixed point of this flow coincides with the
tangency condition and that the corresponding invariant graph exhibit
extremal entropic properties.
\end{abstract}

\maketitle

\section{Introduction}
One of the common transitions between regular and chaotic behavior is
intermittency, the seemingly random alternation of long quasi-regular or
laminar phases, so-called intermissions, and relatively short irregular or
chaotic bursts. Intermittency is omnipresent in nonlinear science and has
been weighed against comparable phenomena in nature, such as
Belousov-Zhabotinski chemical reactions, Rayleigh-Benard instabilities,
turbulence, etc. \cite{schuster, maurer, pomeau, berge}. The study and
characterization of the onset mechanisms and main statistical properties of
intermittency was carried out already a long time ago; Pomeau and Manneville
\cite{pom-mann} introduced a classification as types I-III for different
kinds of intermittency. Subsequently, other types have been described and
typified, such as on-off intermittency \cite{platt}, ring intermittency \cite%
{hramov}, etc. Our objective here is to generate networks from the time
series associated with intermittency and look how this phenomenon translates
into such a different setting, and then examine the manifestation of its
properties in the new context. For definiteness we chose the case of type I
intermittency as it occurs just preceding an (inverse) tangent bifurcation
in nonlinear iterated maps, although the very same methodology can be extended to other situations. Specifically, we show how this phenomenon can be
visualized through the graphs generated when the Horizontal Visibility (HV)
algorithm \cite{lacasaetal2008,luqueetal2009} is applied to
the trajectories of the universality class of unimodal maps, as represented
by the quadratic logistic map.\\

The idea of mapping time series into graphs is actively developed
at present via different approaches \cite{lacasaetal2008, luqueetal2009,
zhang06, kyriakopoulos07, xu08, donner10, donner11, donner11-2, campanharo11}.
Amongst them, the HV approach offers a promising new method for performing
time series analysis,
most of all because it has been corroborated that the fundamental nature of
rather different complex dynamical processes is inherited by the associated
visibility graphs. As part of the effort of
developing a mathematically-sound visibility graph theory of dynamical
systems, in recent years the performance of the visibility method has been
tested and found to be consistently capable in different circumstances,
including the description of chaotic, fractal-stochastic, or dissipative
processes, to cite some \cite{toral10, nunhez12}. In every case the
network counterpart of each particular kind of dynamics has been determined
with precision, positioning the visibility algorithm as a well-defined
method to analyze the dynamics of complex systems using graph-theoretical
tools. In the context of low-dimensional chaos, two main routes to chaos
have been studied in the light of this technique. Specifically, the
period-doubling bifurcation cascade (Feigenbaum scenario) and the
quasiperiodic route have been analyzed through the HV formalism and two
complete sets of graphs, called Feigenbaum and quasiperiodic
graphs respectively, that encode the dynamics of their corresponding classes of iterated
maps, have been introduced and characterized recently \cite{plos, caos, quasi, pla}.
The third well-known route to chaos present in low-dimensional dissipative
systems is type-I intermittency (Pomeau-Manneville scenario), and in the
present work we present the structural, scaling and entropic properties of
the graphs obtained when the HV formalism is applied to this situation.\\

In the following we first
recall in section II the key aspects of type I intermittency.
In section III we present the Horizontal Visibility algorithm and apply it to the study of trajectories generated by unimodal maps close to an inverse tangent bifurcation, where type-I intermittency takes place. A phenomenological
derivation of the degree distribution $P(k;\epsilon)$ of this kind of graphs is performed.
We show that this single graph metric encodes the key scaling properties of
type I intermittency, namely \emph{(i)} the mean length $\langle \ell \rangle $ of the laminar episodes with $%
\epsilon $ manifests in network space as a comparable scaling with the same
variable of the second moment $\langle k^{2}\rangle $ of the degree
distribution $P(k,\epsilon )$, and \emph{(ii)} the scaling of Lyapunov exponent $\lambda(\epsilon)$ is
recovered in network space from the Shannon block entropies over $P(k;\epsilon)$.\\
Next, in section IV we recast the family of HV graphs generated by intermittent series into a
graph-theoretical Renormalization Group (RG) framework and determine the RG
flows close to and at tangency. We show that there are two trivial fixed
points akin to the high and low temperature fixed points in thermal phase
transitions together with a nontrivial fixed point associated with the
tangency condition. Finally, we also determine the extremal entropic properties of
the RG fixed points as well as the entropy evolution along the RG flows.

\begin{figure*}
\centering
\includegraphics[width=0.9\columnwidth]{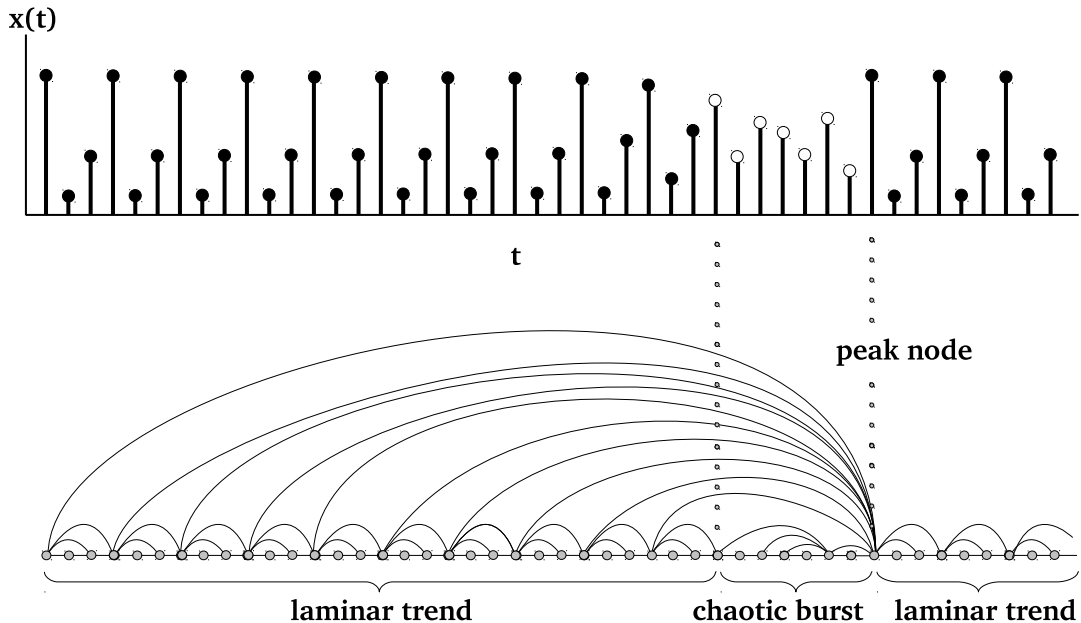}
\caption{Graphical illustration of how the Horizontal Visibility (HV) graph inherits
 in its structure the dynamics of the associated intermittent series.
In the top of the figure we show a sample intermittent series generated
by the logistic map close to $\mu_c$ ($\epsilon>0$), producing laminar regions
(black) mixed with chaotic bursts (white). In the bottom we plot the
associated HV graph.
Laminar regions are mapped into nodes with a periodic backbone,
whereas the actual pseudoperiodicity of the series is inherited in
the graph by the existence of so called peak or interfacial nodes.
Chaotic bursts are mapped into chaotic nodes, with a characteristic
degree distribution (see the text).}
\label{cartoon}
\end{figure*}

\section{Type-I intermittency}

Type-I intermittency can be observed infinitely-many times in the logistic
map
\begin{equation}
x_{t+1}=F(x_t)=\mu x_t(1-x_t),\;0\leq x\leq 1,\;0\leq \mu \leq 4,
\end{equation}%
close to the control parameter values $\mu =\mu_{T}$ at which windows of
periodicity open with period $T$ for values $\mu >\mu _{\infty }\simeq
3.569945672...$., where $\mu_{\infty }$ is the accumulation point of the
main period-doubling cascade that locates the first appearance of chaos when
increasing $\mu$ from small values. It can be observed that at $\mu _{3}=1+\sqrt{8}$ this map exhibits a cycle of period $T=3$ with
subsequent bifurcations. This is the most visible window of periodicity in the chaotic
regime and the one in whose vicinity our simulations have been performed. The regular periodic orbits hold slightly above $\mu_{T}$ but below
$\mu _{T}$ the dynamics consists of laminar episodes interrupted by chaos, a phenomenon known as intermittency.
In what follows we relabel $\mu _{T}\equiv \mu _{c}$ and define $\epsilon \equiv \mu_c-\mu$.
In the upper part of figure \ref{cartoon} we show a sample type-I intermittent time series generated by
the logistic map close to $\mu_3$, showing alternation between laminar trends, represented by
black dots in the series, and chaotic bursts, represented in turn by white dots.
We note that the laminar phase is not actually periodic, but approaches a periodic behavior
of period $3$ and it is precisely this behavior which is close to periodicity
what makes it easily distinguishable from the chaotic bursts.\\
\subsection{Basic properties of intermittent series generated by unimodal maps}
\begin{figure*}
\centering
\includegraphics[width=0.9\columnwidth]{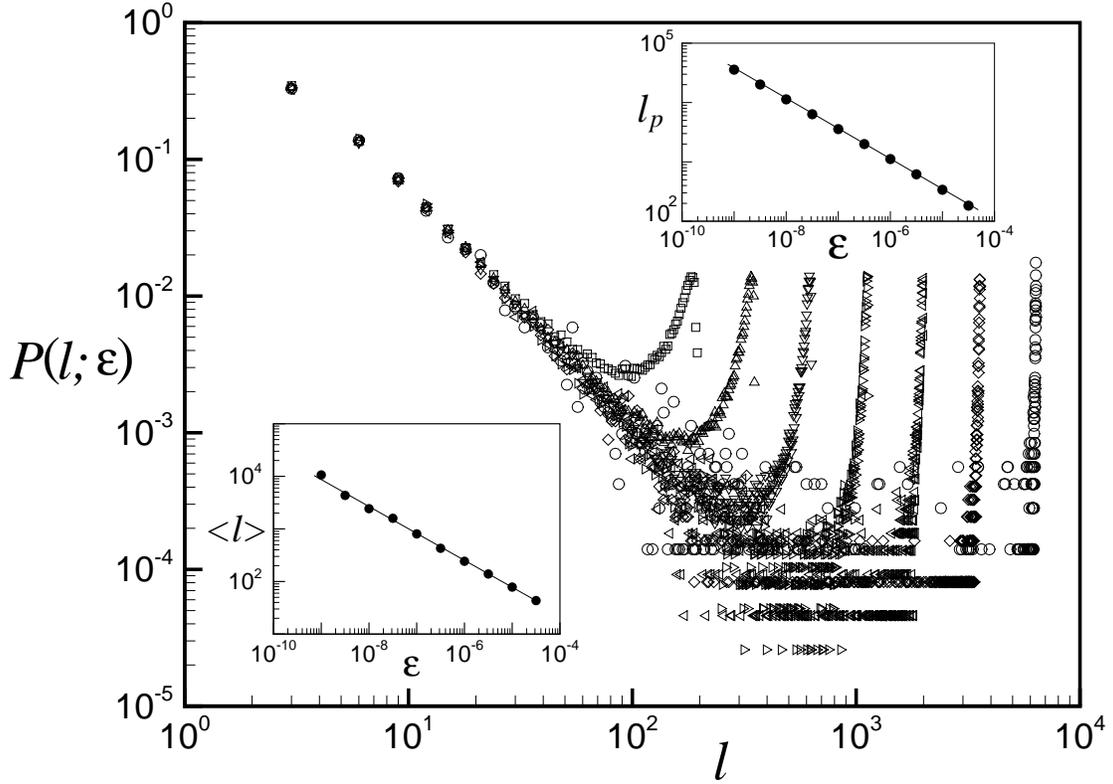}
\caption{Log-log plot of the length distribution of laminar
phases $P(\ell; \epsilon)$ in our system, for different values
of $\epsilon$: (squares) $\protect\epsilon=3\cdot 10^{-5}$, (up-triangles) $\protect\epsilon =10^{-5}$, (down-triangles) $\protect\epsilon =3\cdot10^{-6}$, (right-triangles) $\protect\epsilon =10^{-6}$, (left-triangles) $\protect\epsilon =3\cdot
10^{-7}$, (diamonds) $\protect\epsilon =10^{-7}$, (circles) $\protect\epsilon =3\cdot10^{-8}$. The starting power law decay is saturated at certain
lengths, obtaining the classical asymmetrical U-shaped curves.
(Inset bottom panel) Log-log plot of the laminar phases mean length
$\langle \ell \rangle$ as a function of $\epsilon$, yielding
the expected scaling $\langle \ell \rangle\sim \epsilon^{-0.5}$.
(Inset upper panel) Log-log plot of the local maximum of the distribution $l_p$ as a function of $\epsilon$, also yielding a scaling $l_p\sim \epsilon^{-0.5}$.}
\label{p_trend}
\end{figure*}

Under rather general circumstances, trajectories generated by canonical models evidencing type-I intermittency
show power-law scaling in the mean length of laminar phases
$\langle \ell \rangle \sim \epsilon^{-\gamma}$, where $\epsilon$, called the
channel width of the Poincar\'e section, is the distance between the local
Poincar\'e map and the diagonal \cite{kim}. The specific value of exponent
$\gamma$ is typically associated to the reinjection mechanism and several
exponents have been reported, although $\gamma=0.5$ holds in a rather large set of situations \cite{note1}. In figure \ref{p_trend} we plot
in log-log scales the size distribution of laminar phases
$P(\ell;\epsilon)$, derived numerically from time series of $10^7$ data generated through
the logistic map, for
different values of $\epsilon$, showing the characteristic asymmetric
U-shape. In the bottom inset panel of the same figure, we plot the
dependence of the mean length of laminar trends with $\epsilon$,
showing the well studied scaling $\langle \ell \rangle \sim \epsilon^{-1/2}$.
We finally note that the maximum length of the laminar trends $\ell_p$,
the extremum of the distribution $P(\ell;\epsilon)$,
scales also as  $\ell_p \sim \epsilon^{-1/2}$.\\

\begin{figure*}
\centering
\includegraphics[width=0.9\columnwidth]{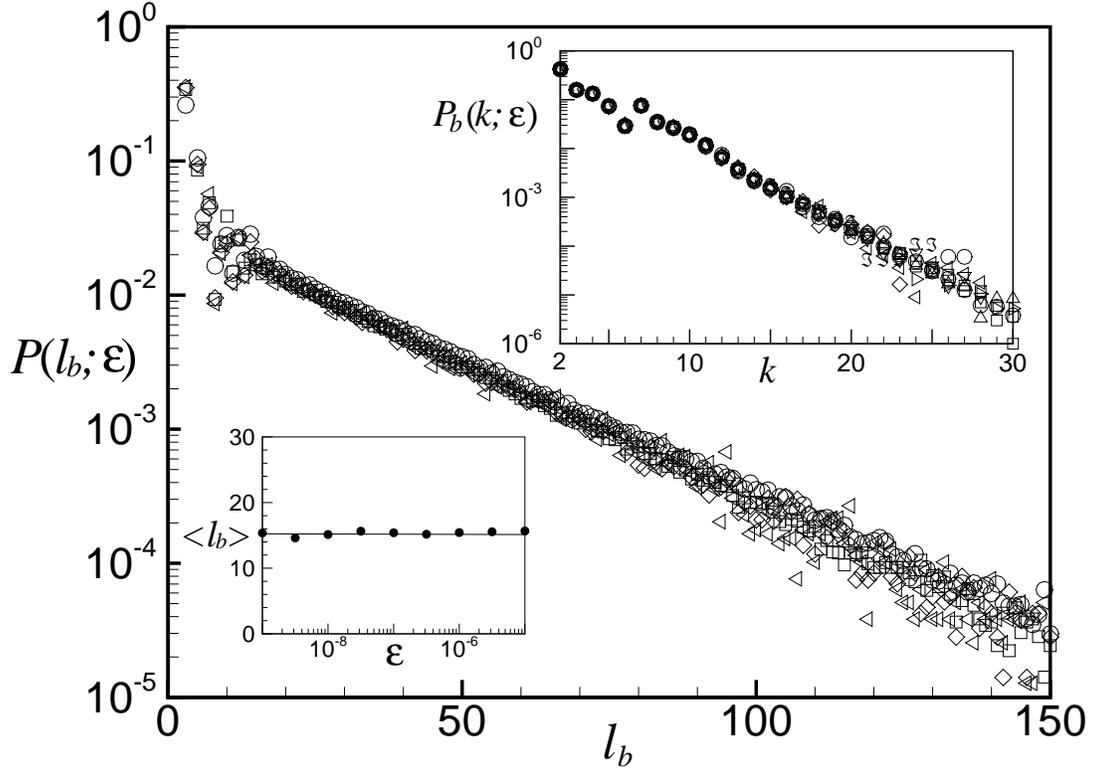}
\caption{Semi-log plot of the size distribution of chaotic bursts in the system,
showing a clear exponential decay which is fairly independent of $\epsilon$.
This behavior, which can be understood as a result of the survival
time of an random uncorrelated process, suggests that the
relaminarization process occurs at unpredictable, uncorrelated
times. (Inset bottom panel) Mean length of the chaotic bursts $\langle l_b \rangle$ for different values of $\epsilon$, showing a constant value $\langle l_b \rangle\approx 15$. (Inset upper panel) Semi-log plot of the degree distribution associated to the
nodes belonging to chaotic bursts $P_b(k)$, for different values of $\epsilon$.
The exponential decay, characteristic of an HV graph associated
to a chaotic series, has been already found in several other maps
\cite{toral10, luqueetal2009}. The results seem to be independent of $\epsilon$,
in good agreement with the fact that the relaminarization process
occurs at unpredictable times, what suggests $P_b(k)=A\exp(-c k)$.}
\label{pburst}
\end{figure*}

On the other hand, the length of a chaotic burst is known to be unpredictable.
In figure \ref{pburst} we plot, in semi-log scales, the size distribution of
chaotic bursts
$P(\ell_b; \epsilon)$ for the same time series as used in figure \ref{p_trend}.
This distribution has in turn an exponential decay which becomes fairly
independent of $\epsilon$ for sufficiently large lengths, a result that can
be justified invoking the survival time of a random memoryless process.
In the lower inset of figure \ref{pburst} we numerically check that,
as $P(\ell_b;\epsilon) \approx P(\ell_b)$, the mean length of the chaotic bursts
remain constant independently of $\epsilon$, with an approximated value of
$\langle \ell_b \rangle \approx 15$.\\

\section{Transformation of intermittent time series into Horizontal Visibility graphs}

The Horizontal Visibility (HV) algorithm \cite{lacasaetal2008, luqueetal2009} assigns
each datum $x_{i}$ of a time series $%
\{x_{i}\}_{i=1,2,...}$ to a node $i$ in its associated HV graph (HVg), where $i
$ and $j$ are two connected nodes if $x_{i},x_{j}>x_{n}$ for all $n$ such
that $i<n<j$. Structural properties of a time series are inherited by its
HVg (see the appendix for some specific properties relevant to the intermittent structure). In the bottom part of figure \ref{cartoon} we show the HV
graph of the associated intermittent series, which consists of several repetitions of
a 3-node motif (periodic backbone) linked to the first node of the subsequent laminar trend, interwoven
with groups of nodes irregularly
(chaotically) connected amongst them.
We observe that the motif repetitions
in the graph correspond to the laminar regions in the trajectory (pseudoperiodic
data with pseudoperiod 3) and the
chaotically connected groups correspond to the chaotic bursts in the
trajectory. As laminar trends are indeed pseudoperiodic in the sense that they
can be decomposed as a periodic signal and a drift (see appendix), this pseudoperiodicity expresses
in the graph structure by allowing a node for each period-3 motif to be connected to
the first node in the next laminar region (the so called peak or interfacial node),
as the values of the time series
in the chaotic bursts are always smaller than those in the former laminar trend. The sequence of
degrees is of the form $2-3-6$ for laminar trends and loses this pattern in the chaotic
burst. At odds with standard approaches, for which the distinction between laminar
and chaotic phases is somewhat ambiguous, in this work we take advantage of this
characteristic pattern as the criterion  to numerically distinguish between both phases.\\
%As a whole, this direct mapping of the intermittent structure in network space
%will allow the identification of characteristic scalings, as will be shown in the next sections.

%\section{Degree distribution of intermittent graphs: characteristic scalings}
\begin{figure*}[tbp]
\centering
\includegraphics[width=0.9\columnwidth]{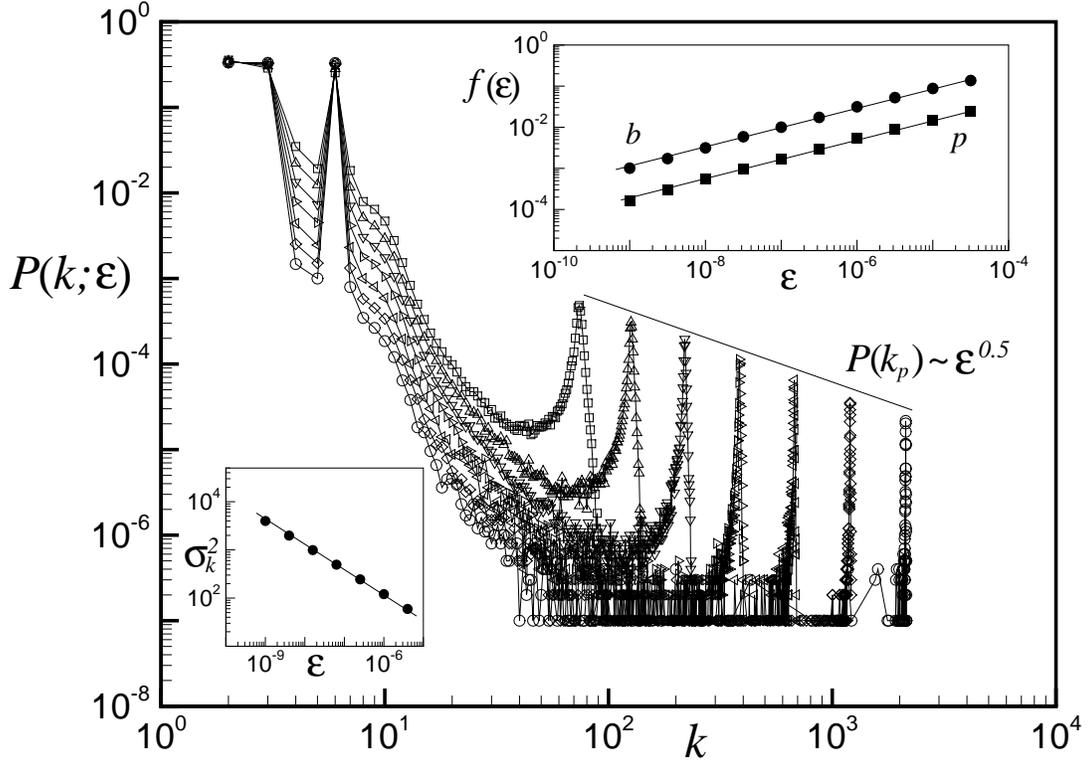}
\caption{Log-log plot of the total degree distribution $P(k;\protect\epsilon )$ of 7 HV graphs mapped
from trajectories of the logistic map in the vicinity of the window of
period 3 ($\epsilon>0$). Squares: $\protect\epsilon=3\cdot 10^{-5}$, Up-triangles: $\protect\epsilon =10^{-5}$, Down-triangles: $\protect\epsilon =3\cdot10^{-6}$, right-triangles: $\protect\epsilon =10^{-6}$, left-triangles: $\protect\epsilon =3\cdot
10^{-7}$, diamonds: $\protect\epsilon =10^{-7}$, circles: $\protect\epsilon =3\cdot10^{-8}$. Note that the tail of each distribution, associated to the peak nodes, scales with $\epsilon^{0.5}$, while this scaling is absent in the tail of the peak degree distribution (see the main panel of figure \ref{p_k_pick}). This scaling is reminiscent of the contribution of the amount of peak nodes present in each series, whose abundance scales with $\epsilon^{0.5}$ (see the inset upper panel of this figure). (Inset upper panel) Log-log plot of the fraction of nodes associated with chaotic bursts, showing a scaling $f_{b}(\epsilon)\simeq 54\cdot \protect\epsilon ^{0.5}$ (circles), and fractions of peak (interfacial) nodes, showing a scaling $f_{p}(\epsilon)\simeq 3.2\cdot \protect\epsilon ^{0.5}$ (squares). (Inset bottom panel) Log-log plot of the
variance of the total degree distribution $\protect\sigma _{k}^{2}=\langle k^{2}\rangle -\langle k\rangle ^{2}
$ as a function of $\protect\epsilon $, obtained from the
same graphs whose degree distribution is plotted in the main panel. A power law scaling of the form $\protect\sigma _{k}^{2}\sim \protect%
\epsilon ^{-0.5}$ is found, which is the graph analogue of the well-known laminar phase scaling reported in the inset bottom panel of figure \ref{p_trend}.}
\label{p_k}
\end{figure*}

\begin{figure*}[tbp]
\centering
\includegraphics[width=0.9\columnwidth]{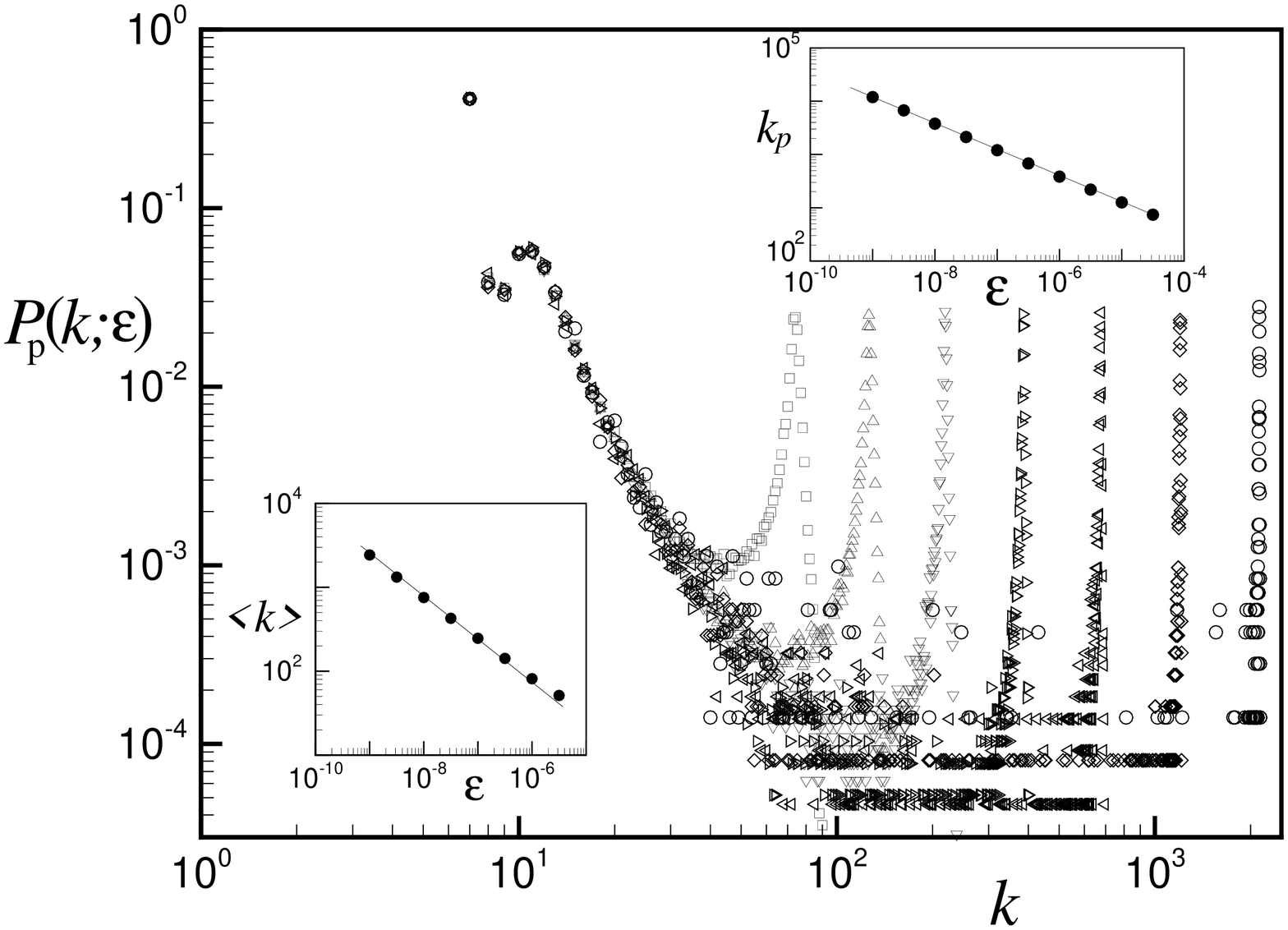}
\caption{Log-log plot of the degree distribution of the so called peak nodes, obtained from the HV graphs at different values of $\epsilon$. The asymmetrical U-shape is qualitatively similar to the size distribution of laminar phases (see the text). (Inset bottom panel) Log-log plot of the mean degree of peak nodes, as a function of $\epsilon$, calculated from the distributions plotted in the main panel of the figure. The scaling coincides with $\langle \ell \rangle/3$ (see the text). (Inset upper panel) Log-log plot of the local maximum of the distribution $k_p$ as a function of $\epsilon$ showing a scaling $k_p\sim \epsilon^{-0.5}$.}
\label{p_k_pick}
\end{figure*}

The assembly of repeated 3-node motifs separated by sets of nodes with
chaotic links inherited by the HVg from the laminar trends and chaotic
bursts in the intermittent series, leaves also a characteristic footprint in
its degree distribution $P(k;\epsilon )$ (see figure \ref{p_k}). This can be
seen when $P(k;\epsilon )$ is compared with that for fully chaotic motion
(see for instance \cite{toral10}). In the former the connectivity of the nodes in the
3-node motifs is overrepresented than in the latter, since their
relative frequencies are proportional to the length of the laminar
episodes. Also, nodes with large degree (peak or interfacial nodes), increasingly large as $\epsilon$
decreases, emerge due to reinjections after chaotic bursts because these
have visibility over the nodes from laminar phases. The evidence
collected leads us to express the total degree distribution
$P(k;\epsilon)$ as composed of three contributions that originate from three
different types of nodes. Namely:\\

\textbf{i) Laminar:} the contribution from the laminar phases, $P_{l}(k;\epsilon )$, consists of a discrete set of
degrees that correspond to the background periodic behavior. By construction (see appendix)
a periodic series with a superimposed drift generates a graph where the
nodes have, for the particular case in which $T=3$, a degree belonging to the set $\{2,3,6\}$, therefore $P_{l}(k;\epsilon )\equiv P_l(k)=1/3$ when $k=2,3,6$, and zero otherwise.\\

\textbf{ii) Chaotic:} the contribution from the chaotic bursts, $P_{b}(k;\epsilon )$,
which, according to previous works \cite{luqueetal2009, toral10}, has an exponential decay (see the upper inset panel of figure \ref{pburst} for numerical
evidence). Moreover, as argued in the previous section, since the mean size of chaotic bursts is independent of $\epsilon$ (as we can see in the
inset bottom panel of figure \ref{pburst}), the contribution $P_{b}(k;\epsilon )$ is thus also independent of $\epsilon$, and in general reads
$P_{b}(k;\epsilon)= P_b(k)\sim \exp(-c k)$, where the specific value of $c$ depends on the chaos dimensionality \cite{toral10}.\\

\textbf{iii) Peak:} the contribution from the interface between the chaotic and
laminar phases, $P_{p}(k;\epsilon )$ (see figure \ref{p_k_pick}), arises from the peak nodes, with a very large degree that is approximately
proportional to the size of the laminar phase. Roughly, all of the peak nodes inherit a degree based on their visibility of the previous laminar phase,
and therefore we a priori may assume that $P_{p}(k;\epsilon)$ and $P(\ell/3 ;\epsilon)$ (see figures \ref{p_trend} and \ref{p_k_pick}).
The factor of $1/3$ arises because, for the periodic window studied, $k \sim \ell /3$ (or $k \sim \ell /T$ if the study focuses on intermittency close to a periodic window of period $T$), as we can observe in figure \ref{cartoon}. In particular, the maximum laminar size and peak node degree scale similarly, $ \ell_p/3 \approx k_p\sim \epsilon^{-0.5}$ (see the upper panels of figures \ref{p_trend} and \ref{p_k_pick}). Note however that the connectivity $k$ of the peak nodes
is not straightforwardly distributed as $\ell$, since the actual value of
the degree of the peak node assigned by the HV is not necessarily equal to
the size of the full laminar phase, but varies in somewhat according to the
actual value (position) of the series datum at reinjection (compare figures \ref{p_k_pick} and \ref{p_trend}). That is to say,
visibility of a peak node may be as large as the preceding laminar trend, but it can also be smaller if reinjection takes place
below tangency, or larger, allowing full visibility of the preceding trend and part of the previous one, if reinjection takes
place above tangency.\\

Nevertheless, in the following we will argue, and have numerically checked, that $P_{p}(k;\epsilon)$ and $P(\ell/3 ;\epsilon )$ have, up to first order, similar first and second moments, concretely: (a) The mean degree of peak
nodes is approximately equal to $1/3$-rd the mean value of the laminar phase size. This is due to the fact that
the degree of peak nodes statistically \emph{self-averages} over laminar phase sizes, as the variability in reinjection
is symmetrical with respect to tangency. And (b) the second moment of $P_{p}(k;\epsilon )$ is associated with both the
second moment of $P(\ell ;\epsilon )$ and the variance of the reinjection distribution.
However, the contribution of the latter is typically much smaller than the former. Indeed, whereas fluctuations in the
reinjection position tend to decrease as $\epsilon$ decreases, fluctuations in the size of laminar trends tend to
increase as $\epsilon$ decreases (this has been confirmed numerically). Therefore, the leading contribution comes
from the variability of laminar sizes for small values of $\epsilon$.\\

The aforementioned phenomenology lead us to formally write down $P(k;\epsilon)$ as
\begin{equation}
P(k;\epsilon )=f_{l}(\epsilon)\cdot P_{l}(k)+f_{b}(\epsilon)\cdot P_{b}(k)+f_{p}(\epsilon)\cdot P_{p}(k;\epsilon ),
\label{pp}
\end{equation}%
where $f_{l}(\epsilon),f_{b}(\epsilon)$ and $f_{p}(\epsilon)$ are the fractions of nodes in the graph that
correspond to laminar, chaotic and peak regions respectively.
%In what follows we proceed step by step to explicitly calculate each of the terms in equation \ref{pp}. \\
In order to derive these fractions, we rely on three restrictions, namely:\\

\noindent (i) \emph{Normalization:} which trivially implies $f_l(\epsilon)+f_b(\epsilon)+f_p(\epsilon)=1$,\\

\noindent (ii) \emph{Bounded degree:} it has been proved \cite{plos,caos,luqueetal2009} that aperiodic
series generate an HVg with constant mean degree $\langle k\rangle =4$, the upper bound value for HV graphs.
This restriction implies
 \begin{equation}
 f_{l}(\epsilon)\cdot\sum_{k}kP_{l}(k)+f_{b}(\epsilon)\cdot\sum_{k}kP_{b}(k)+f_{p}(\epsilon)\cdot\sum_{k}kP_{p}(k ; \epsilon )=4.
 \end{equation}
 Note that in the latter expression, the first sum is trivially $\sum_{k}kP_{l}(k)=11/3$, the second sum
corresponds to the graph associated with a chaotic series, and is directly $\sum_{k}kP_{b}(k)=4$ \cite{plos,caos},
while the third sum yields, due to the aforementioned arguments, $\sum_{k}kP_{p}(k;\epsilon )\approx \langle\ell \rangle/3$.\\

\noindent (iii) After each chaotic burst a peak node emerges and anticipates the next laminar region,
what implies $f_b=f_p \cdot \langle \ell_{\text{b}}\rangle$, where $\langle \ell_{\text{b}}\rangle$
is the mean size of a chaotic burst and has been argued to be $\epsilon$-independent.\\

 After a little algebra, (i), (ii) and (iii) along with the rest of the arguments yield the prediction $\langle \ell_{\text{b}}\rangle=11$,
 $f_b = 11 \cdot f_p$, $f_l=1-12 \cdot f_p$ and $f_p=\langle \ell\rangle^{-1}$.

In order to compare our phenomenological prediction with the numerics, we recall that under rather general conditions
$\langle\ell \rangle \approx 0.2 \cdot \epsilon^{-0.5}$ \cite{note2}, which indeed coincides with the
scaling plotted in the lower inset panel of figure \ref{p_trend}. The predicted values for the
fractions are therefore $f_p=5 \cdot \epsilon^{0.5}, f_b=55\cdot \epsilon^{0.5}$,
and $f_l=1-60\cdot \epsilon^{0.5}$. These can be compared with the results of numerical
simulations, shown in the upper inset panel of figure \ref{p_trend}, for which the best fit
are $f_{p}\approx 3.2 \cdot \epsilon ^{0.5 }$, $f_{b}\approx 54 \cdot \epsilon ^{0.5}$, $f_{l} \approx 1- 57.2 \cdot \epsilon ^{0.5}$, and additionally $\langle \ell_{\text{b}}\rangle\approx15$ (bottom inset panel of figure \ref{pburst}),
on fairly good agreement with our prediction.\\

\subsection{Variance $\protect\sigma _{k}^{2}=\langle k^{2}\rangle
-\langle k\rangle ^{2}$: graph analogue of $\langle \ell \rangle$}

%The preliminary visual observation in figure \ref{p_k} of the scaling $k_{peak}\sim
%\epsilon ^{-\alpha}$ is made quantitative in the up inset right panel of figure \ref{p_k_pick} where it can be
%seen that the local maximum in $P(k;\epsilon )$ obeys the same power law ($\alpha=0.5$) as
%the mean length of the laminar trends}.
While there can be no equivalence between $\langle \ell \rangle $ and $\langle k\rangle $ in the intermittent graphs as
the latter is fixed to be $\langle k\rangle =4$ for an aperiodic regime \cite{nunhez12}, a relationship may hold between $\langle \ell \rangle $\
and higher moments of $P(k;\epsilon )$.
%to explain the apparent {\bf relation
%between $\langle \ell \rangle$ and $\langle k \rangle$ for the peak node distribution.
Note that there is an increasing
dispersion of the values of $k$ from its mean $\langle k\rangle $ in the
degree distributions of the graphs as $\epsilon \rightarrow 0^{+}$, that we
show below is related to the distribution connectivity $P_p(k;\epsilon)$. If we measure this dispersion by means
of the variance of the total degree distribution $\sigma _{k}^{2}=\langle
k^{2}\rangle -{\langle k\rangle }^{2}$  we
recover numerically the $\epsilon ^{-0.5}$ scaling, as shown in the bottom inset panel of figure \ref{p_k}.\\

This scaling is also a prediction of our phenomenological theory:
\begin{eqnarray}
\sigma _{k}^{2} &=&\langle k^{2}\rangle -\langle k\rangle ^{2}  \nonumber \\
&=&f_l(\epsilon)\sum_{k}k^{2}P_{l}(k)+f_b(\epsilon)\sum_{k}k^{2}P_{b}(k)+f_p(\epsilon)\sum_{k}k^{2}P_{p}(k;\epsilon)-\langle
k\rangle ^{2}.
\end{eqnarray}%
Proceeding as before, the first sum is $\sum_{k}k^{2}P_{l}(k)=49/3$ and the second sum $\sum_{k}k^{2}P_{b}(k)$ is finite and $\epsilon$-independent. To determine the third sum we
recall that the variance the distributions $P_p(k;\epsilon)$ and $P(\ell;\epsilon)$ are equal up to first order in $\epsilon^{-1}$ (the variance associated with the reinjection probability is always a smaller quantity), yielding $\sum_{k}k^{2}P_{p}(k;\epsilon )\sim \sum_{\ell }\ell ^{2}P(\ell
;\epsilon )\sim
\epsilon ^{-1}$, where the higher-order terms
take into account the fluctuations associated to the reinjection variability. Collecting these results we straightforwardly find
\begin{equation}
\sigma _{k}^{2}\sim \epsilon ^{-0.5},
\end{equation}
for small values of $\epsilon$. This is a main quantitative result linking the key
property of intermittent time series with its counterpart in the
corresponding HV graphs.

\begin{figure}
\centering
\includegraphics[width=0.55\columnwidth]{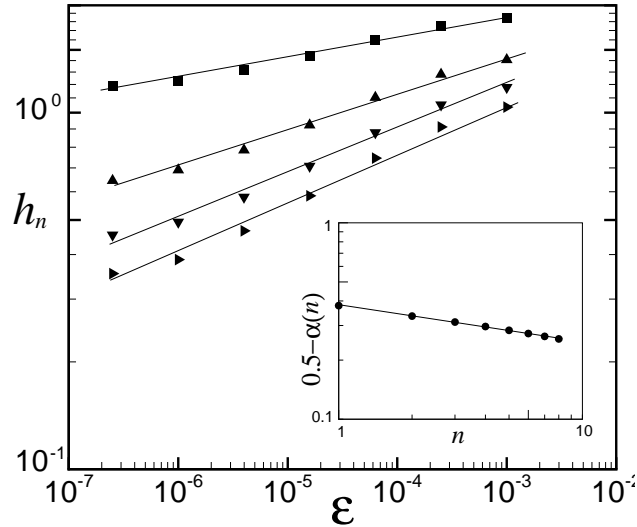}
\caption{Log-log plot of the block entropies $h_{n}$ constructed from degree distributions of $n$-sequence of connectivities in the HV graphs
as a function of $\epsilon$: $n=1$ (squares), $n=2$ (up-triangles), $n=3$ (down-triangles), $n=4$ (right-triangles). A scaling of the form $h_{n} \sim \epsilon^{\alpha(n)}$, reminiscent of the well known scaling of the Lyapunov exponent $\lambda\sim\epsilon^{0.5}$
is found, albeit with a different, $n$ dependent exponent $\alpha(n)$. (Inset panel) Log-log plot of the convergence of $\alpha(n)$ to the exponent associated to the Lyapunov exponent, as a function of $n$.
A relation of the form $[0.5 -\alpha(n)]\sim n^{-0.19}$ is found, whose extrapolation suggests $\lim_{n\rightarrow\infty}\alpha(n)=0.5$}
\label{blk_entr}
\end{figure}

To round off, we here provide as an ansatz a concrete expression for $P_p(k; \epsilon)$:
\begin{equation}
P_p(k; \epsilon) \sim k^{-\alpha} \frac{\epsilon^{\alpha/2}}{|k-k_p|+1},
\label{ajuste}
\end{equation}
where $k_p=a\epsilon^{-0.5}$. This ansatz complies with all the properties obtained in our phenomenological theory (although note that we do not need to use it to derive several observables as the fraction of nodes or the variance $\sigma^2_k$), and provide a closed expression for the total degree distribution. This closed expression allows us to calculate the concrete degree distribution at tangency.
%\begin{equation}
%\lim_{\epsilon\rightarrow 0} P(k;\epsilon) =
%\left\{
%\begin{array}{rcl}
%     P_l(k),\  k=2,3,6
%  \\ \delta(\ell/3-\infty),\  \textrm{otherwise}
%\end{array}
%\right.
%\label{pktotalcero}
%\end{equation}
This consists of an infinite-size laminar phase and a `phantom' of $k_{p}$ with diverging degree and vanishing probability of occurrence. Incidentally, note that
in this limit case we also recover $\langle k\rangle =\sum k P(k;\epsilon\rightarrow 0)=4$, as expected.

\subsection{Scaling of Lyapunov exponent: Block entropies $h_n$}
In the previous section we have studied how the scaling of $\langle l \rangle$ was inherited in $P(k)$ by $k_{p}$ and
$\sigma_k^2$, but there exists another well known scaling relation in the type-I intermittency route involving the Lyapunov exponent
\begin{equation}
\lambda=\lim_{n\rightarrow\infty}\frac{1}{n}\sum_{i=0}^{n-1}\ln|F'(x_{i})|
\end{equation}
of the trajectories \cite{pom-mann,hirsch}, which reads $\lambda\sim \epsilon^{0.5}$ as $\epsilon\rightarrow 0$. We recall here that the Pesin identity relates the positive Lyapunov exponents of a chaotic dynamics with the Kolmogorov-Sinai entropy of
the system. In recent works \cite{plos,caos,pla}, such relation has been investigated in the visibility graph framework, through the definition of a graph theoretical entropy, a Shannon-like entropy over the degree distribution
\begin{equation}
h_1=-\sum_{k}P(k)\log P(k).
\end{equation}
In figure \ref{blk_entr} we plot in log-log the values of $h_1$ (solid squares) as a function of the channel width $\epsilon$. A power law scaling is
recovered, albeit with a different scaling exponent $\alpha<0.5$. Notice however that $h_1$ is only a proxy of the Kolmogorov-Sinai entropy and thus
a comparison with the Lyapunov exponent is only approximate. Interestingly, $h_1$ is indeed the graph theoretical version of a size-1
block entropy over the degree distribution. Since the Kolmogorov-Sinai entropy of a map can be recovered as
the asymptotic limit of block entropies \cite{nicolis} $s(n)=-\frac{1}{n}\sum_{x_{1},...,x_{n}}p(x_{1},...,x_{n})\log p(x_{1},...,x_{n})$, we take
advantage of this fact to define a set of graph-theoretical block entropies
\begin{equation}
h_{n}=-\frac{1}{n}\sum_{k_{1},...,k_{n}}P(k_{1},...,k_{n})\log P(k_{1},...,k_{n}).
\end{equation}
The $\epsilon$-dependence of these block entropies are also plotted in figure \ref{blk_entr}, for different block sizes $n$. We observe in every case a power law scaling $h_n\sim \epsilon^{-\alpha(n)}$. In the inset of the same figure we show how the exponent $\alpha(n)$ converges with $n$ to the exponent found for the scaling of the Lyapunov exponent, suggesting
\begin{equation}
\lim_{n\rightarrow\infty}h_{n}=\lambda.
\label{Pesin}
\end{equation}
 We remark at this point that, whereas the entropy is a magnitude defined in the graph, the Lyapunov exponent is only defined in the system. Therefore, in rigor the Pesin identity cannot be used here as the explanation for equation $\ref{Pesin}$, as we are mixing properties defined in two different contexts. However, the strong numerical evidence in favor of a Pesin-like identity between the map's Lyapunov exponent and the graph's block entropy suggest that a graph analogue of the Lyapunov exponent may be defined in network space \cite{pla}.

\begin{figure*}[tbp]
\centering
\includegraphics[width=0.9\columnwidth]{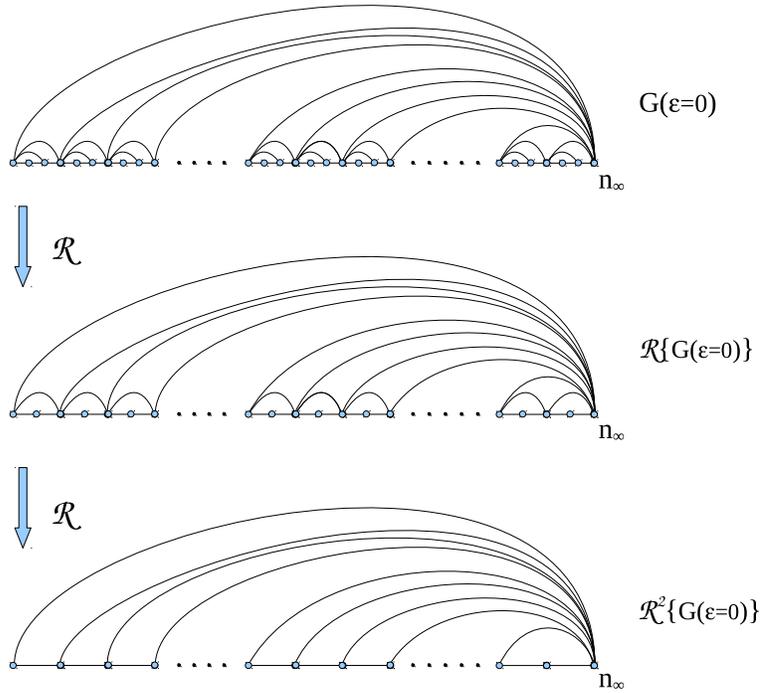}
\caption{Illustration of the renormalization operator $\cal R$ applied on the HV graph at $\epsilon=0$. This graph renormalizes, after two iterations of $\cal R$, into an HV graph $G_c$ which is itself (i) invariant under $\cal R$, and (ii) unstable under perturbations in $\epsilon$, thus constituting a nontrivial (saddle) fixed point of the graph-theoretical RG flow.}
\label{RG2}
\end{figure*}

\section{Graph-theoretical renormalization group analysis}

Once the characteristic scalings of type I intermittency have been recovered in the topology of the asssociated graphs, let us provide a wider picture of the phenomenon. The overall properties of intermittency graphs can be framed in the context
of a Renormalization Group (RG) transformation by following the procedure of
previous studies \cite{plos, caos, quasi}. We define an RG transformation $%
\mathcal{R}$ on an HV graph $G$ as the coarse-graining of every couple of
adjacent nodes where at least one of them has degree $k=2$, into a block node
that inherits the links of the previous two nodes. Iterating $\mathcal{R}$
we can trace the RG flows of intermittent graphs $G(\epsilon)$. Results include:\\

\noindent (i) When $\epsilon <0$ ($\mu
\gtrsim \mu _{c}$) trajectories are periodic and every HV graph trivially renormalizes
towards the so called chain graph $G_{0}$, an infinite chain with $k=2$ for all
nodes \cite{plos, caos}. $G_{0}$ is invariant under
renormalization $\mathcal{R}\{G_{0}\}=G_{0}$, and indeed constitutes a trivial
(attractive) fixed point of the RG flow, $\mathcal{R}^{(n)}\{G(\epsilon <0)\}=G_{0}$.\\

\noindent (ii) When $\epsilon >0$ ($\mu \lesssim \mu _{c}$) repeated RG transformations
eliminate progressively the links in the graph associated with correlations
in the time series, leading ultimately to the HV graph that corresponds to a
random time series. The links that stem from temporal correlated data
connect primarily laminar nodes, whereas the links between either burst and
peak nodes originate from uncorrelated segments of the time series. If the
laminar episodes are eliminated from the time series, the burst and
reinjection data values form a new time series that upon renormalization
leads to the random time series. We have
$\lim_{n\rightarrow \infty} {\cal R}^{(n)}\{G(\epsilon>0)\}=G_{\text{rand}}$
, where $G_{\text{rand}}$ is
the HVg associated with a random uncorrelated process with known graph
properties \cite{caos}. This constitutes the second (attractive) fixed point of the RG flow.\\

\begin{figure}[tbp]
\centering
\includegraphics[width=0.35\columnwidth]{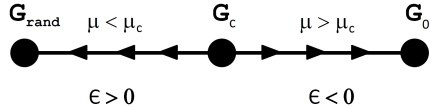}
\caption{Graph-theoretical RG flow of intermittent graphs. Graphs
renormalize to $G_{0}$ for $\protect\epsilon <0$ and to $G_{\text{rand}}$
for $\protect\epsilon >0$, the trivial (attractive) fixed points of the
flow. The graph obtained at $\protect\epsilon =0$ renormalizes into $G_c$, which invariant under the RG
transformation and unstable under perturbations in $\epsilon$, constituting the nontrivial graph-theoretical fixed point of the dynamics, whose
degree distribution coincides with Eq. (\protect\ref{pkRG}) for $\protect%
\epsilon =0$.}
\label{RG}
\end{figure}

\noindent (iii) At $\epsilon =0$ ($\mu =\mu _{c}$) the HV graph
generated by trajectories at tangency converges after only two steps of the
RG transformation to a nontrivial fixed point $\mathcal{R}^{2}\{G(\epsilon
=0)\}=G_{c}=\mathcal{R}\{G_{c}\}$ and remains invariant under $\mathcal{R}$
afterwards. This feature can be demonstrated by explicit application of $%
\mathcal{R}$ upon $G(\epsilon =0)$ (see figure \ref{RG2} for a graphical illustration of this process). The fixed-point graph $G_{c}$ is the HVg
of a monotonically decreasing time series bounded at infinity by a large
value, $G_{c}$ is unstable under perturbations in $\epsilon $ and it is thus
technically a saddle-point of the RG flow, attractive along the critical manifold (spanned by $G(\epsilon=0)$ and its replicas within other periodic windows of period $T$). The RG flow diagram is shown in
Figure \ref{RG}.\\

From visual inspection of figure \ref{RG2} the degree distribution for $G(\epsilon =0)$ is by construction
\begin{equation}
P(k;0)=\lim_{N\rightarrow \infty }\left\{
\begin{array}{rcl}
\frac{N/3}{N} & \text{if} & k=2,3, \\
%\frac{N/3}{N} & \text{if} & k=3, \\
\frac{(N/3)-1}{N} & \text{if} & k=6, \\
\frac{1}{N} & \text{if} & k=N/3,%
\end{array}%
\right.   \label{pkGmuc}
\end{equation}%
and zero otherwise, where $N$ is the number of nodes. Note that this expression coincides with the one predicted from our phenomenological theory when $\epsilon \rightarrow 0$. Its first moment is $\langle k\rangle=4$ while its second moment $\langle k^{2}\rangle$ diverges, as expected from the previous result $\sigma^2_k\sim\epsilon^{-0.5}$. By construction the degree
distribution of $G_{c}$ is
\begin{equation}
P_{G_{c}}(k)=\lim_{N\rightarrow \infty }\left\{
\begin{array}{rcl}
\frac{N-2}{N} & \text{if} & k=3, \\
\frac{1}{N} & \text{if} & k=2, \\
\frac{1}{N} & \text{if} & k=N-1,%
\end{array}%
\right.   \label{pkRG}
\end{equation}%
and zero otherwise. The mean degree is again $\langle k\rangle=4$ and its second moment
also diverges.\\

Let us return now, within our RG treatment, to the concept of graph-theoretical entropies $h_n$.
We first look at the dependence of $h_n$ on $\epsilon$. This is
shown when $\epsilon >0$ in figure \ref{blk_entr} in logarithmic scales and
observe that these functions are power laws that reach in every case a minimum
at tangency. For concreteness we focus on $h_1$, for which $h_1(\epsilon
=0)=\log 3$. This minimum value of $h_1$ is retained for all $\epsilon <0$
(but with $\left\vert \epsilon \right\vert $ below the period-doubling
bifurcations that take place within the window of period three). Hence,
entropy reaches a global minimum for the HV graph at tangency $\epsilon =0$.
Next we enquire about the effect of the RG transformations on $h_1$. The
entropy at the nontrivial fixed point vanishes, as $h_1\left[ P_{G_{c}}(k)%
\right] \rightarrow 0$ when $N\rightarrow \infty $, that is, the RG reduces $%
h_1$ when $\epsilon =0$. Also, the RG transformations increases $h_1$ when $%
\epsilon >0$ (as $h_1[P_{G_{\text{rand}}}(k)]$ $=\log (27/4)$  \cite{plos,caos}) and it reduces it
when $\epsilon <0$ (since $h_1\left[ P_{G_{0}}(k)\right] =0$  \cite{plos,caos}). When $\epsilon
>0$ the renormalization process of removal at each stage of all nodes with $%
k=2$ eliminates node-node correlations (temporal correlations of the
intermittent dynamics) and leads to a limiting renormalized system that
consists only of a collection of uncorrelated variables, generating an
irreversible flow along which the entropy grows. On the other hand, when $%
\epsilon <0$ renormalization increments the fraction of nodes with degree $%
k=2$ at each stage driving the graph structure towards the simple chain $%
G_{0}$ and thus decreases its entropy to its minimum value. Thus we observe
the familiar picture of the RG treatment of a model phase transition, two
trivial fixed points that represent disordered and ordered, or high and low
temperature, phases, and a nontrivial fixed point with scale-invariant
properties that represents the critical point. There is only one relevant
variable, $\epsilon =\mu _{c}-\mu $, that is necessary to vanish to enable
the RG transformation to access the nontrivial fixed point. The property
that is seldom observed \cite{robledo1} is that an entropy functional, in
the present case $h_1[P(k;\epsilon)]$, varies monotonously along the RG flows and is
extremal at the fixed points. A salient feature of the HV studies of the
routes to chaos in low-dimensional nonlinear iterated maps, period doubling
\cite{plos,caos}, quasiperiodicity \cite{quasi}, and intermittency as
presented here, is the demonstration that the entropy functional $h_1[P(k)]$
behaves as mentioned and attains extremal (maxima, minima or saddle-point)
values at the RG fixed points.

\section{Summary}

We have demonstrated the capability of the HV algorithm for transforming
into network language the properties of the route to chaos via intermittency
of type I as it occurs in unimodal one-dimensional iterated maps. The
outcome is a novel type of network architecture composed by nodes of three
types with characteristic connectivities and that are the building blocks
with which intermittency is expressed recursively via concatenation in
network space. These node types arise from the laminar trends, the chaotic
bursts, and the interfacial positions related to reinjection into the
channel. Their relative numbers as a function of the channel width $\epsilon
$ and their contributions to the degree distribution $P(k)$ were determined from a phenomenological theory, with accurate results in agreement with numerical simulations. We have shown that the characteristic ingredients of
type-I intermittency are inherited by the graph as functionals of the latter degree distribution, namely (i)
the graph analogue of the laminar phase mean length, which evidences the well known scaling $\langle \ell \rangle \sim \epsilon^{-0.5}$, was identified to be the variance of the total degree distribution, and (ii) the Shannon block entropy over the degree distribution appears as the graph analogue of the Lyapunov exponent $\lambda$, as the characteristic scaling $\lambda\sim \epsilon^{-0.5}$
is asymptotically recovered in network space through graph-theoretical block entropies $h_n$. We note at this point that Pesin identity suggests that a graph theoretical analogue to the Lyapunov exponents could be defined in the network context \cite{pla}. We also would like to highlight that while we have focused, for definiteness, in the logistic map close to the periodic window of period 3, results can be trivially extended to the intermittent dynamics close to any given periodic window or any chaotic map undergoing an inverse tangent bifurcation.\\

Significantly, the
HV formalism leads to analytical expressions for the degree distribution
near and at tangency. The scaling properties of the intermittent networks
can be determined in terms of the same RG transformation employed with
success on the HV graphs obtained for the period doubling and
quasiperiodicity routes to chaos \cite{plos, caos, quasi}. The graph-theoretical RG
fixed points capture the features of the dynamics above, below and at the
tangent bifurcation. Finally, the optimization of a graph entropy
introduced via the degree distribution reproduces the RG flows and fixed
points.\\

In conclusion, the transition to chaos via type-I
intermittency, as exemplified in unimodal maps near and at an inverse tangent
bifurcation, has been fully described within the Horizontal Visibility theory.
This technique may be extended to the study of other types of
intermittency and may be useful for the analysis and interpretation of time
series with sporadic features of diverse origin.\\

\textbf{Acknowledgements.} We acknowledge financial support by the MEC and
Comunidad de Madrid (Spain) through Project Nos. FIS2009-13690 and
S2009ESP-1691 (A.N., B.L. and L.L.), and support from CONACyT CB-2011-01-167978 \& DGAPA (PAPIIT
IN100311)-UNAM (Mexican agencies) (A.R.).

\section{Appendix}

Some properties of HV graphs which are relevant to the characteristics of intermittent series include:\\

i) \emph{Drift}: Let $\{x_{i}\}_{i=1,2,...}$ be a monotonically
increasing/decreasing series and define its drift as $\{d_{i}\}_{i=1,2,...}%
\equiv $ $\{x_{i+1}-x_{i}\}_{i=1,2,...}$. Every node $i$ of its HVg is
connected only to the previous node $i-1$ and to the following node $i+1$,
as it is always true that $x_{i+1}\geqslant x_{i}\geqslant x_{i-1}$ or $%
x_{i+1}\leqslant x_{i}\leqslant x_{i-1}$. Therefore every node has
connectivity $k=2$ and the degree distribution is $P(k=2)=1$.\\

ii) \emph{Periodicity:} Let $\{x_{i}\}_{i=1,2,...}$ be a periodic series of period $%
T$ so that $x_{i+T}=x_{i}$. Let $x_{max}$ be $\max\{x_{k}\}_{k=i,...,i+T}$. Its
HVg consists of a repeated periodic motif of $T$ nodes from node $i_{max}$
corresponding to $x_{max}$ to node $i_{max}+T$ corresponding to $%
x_{max+T}=x_{max} $ \cite{nunhez12}.\\

iii) \emph{Periodicity with drift:} Let $\{x_{i}^{d}\}_{i=1,2,...}$ be a drifted periodic
series with drifted period $T$ so that $x_{i+T}^{d}=x_{i}^{d}+d_i$. If the series satisfies $\mid
x_{i+1}^{d}-x_{i}^{d}\mid >\mid d_{i}-d_{i+1}\mid $ for all$\;i$, the HV graphs
associated to $\{x_{i}^{d}\}$ and to $\{x_{i}\}$ in ii) are identical.\\

iv) \emph{Chaoticity:} Let $\{x_{i}\}_{i=1,2,...}$ be a chaotic series. The degree
distribution of its HVg has an exponential tail $P(k)\sim \exp (-\lambda k)$
with $\lambda \geqslant \ln (3/2)$, the specific value of $\lambda $ depends
on the chaotic process from which the series has been extracted from \cite{toral10}.\\

 Also note that the initial period-3
cycle behaves as $x_{2}=\mu _{c}x_{1}(1-x_{1})$, $x_{3}=\mu
_{c}x_{2}(1-x_{2})$, $x_{1}=\mu _{c}x_{3}(1-x_{3})$, and in the chaotic
vicinity of this orbit we have
\begin{eqnarray}
x_{2}^{d} &=&(\mu _{c}-\epsilon )x_{1}(1-x_{1})=\mu
_{c}x_{1}(1-x_{1})-\epsilon \lbrack x_{1}(1-x_{1})]=x_{2}-\epsilon \mu
_{c}^{-1}x_{2}, \\
x_{3}^{d} &=&(\mu _{c}-\epsilon )x_{2}(1-\epsilon \mu
_{c}^{-1})[1-x_{2}(1-\epsilon \mu _{c}^{-1})]\approx x_{3}-\epsilon \mu
_{c}^{-1}(2x_{3}-\mu _{c}x_{2}^{2}), \\
x_{1}^{d} &\approx &x_{1}-\epsilon \mu _{c}^{-1}(x_{1}-4\mu
_{c}x_{3}^{2}+2\mu _{c}x_{3}+2\mu _{c}^{2}x_{3}x_{2}^{2}-\mu
_{c}^{2}x_{2}^{2})
\end{eqnarray}%
that can be seen to be a perturbed periodic orbit. If we consider the series
$\{x_{i}^{d}\}_{i=1,2,...}$ to be periodic with a drift, $x_{i+T}^{d}=x_{i}^{d}+d_i$,
then according to condition (iii) above the
periodic HVg remains invariant for pseudoperiodic orbits $\{x_{i}^{d}\}_{i=1,2,...}$.

\bibliography{apssamp}% Produces the bibliography via BibTeX.

\begin{thebibliography}{99}
\bibitem{note1}
For controlled reinjections, this scaling is of the form $%
\langle \ell \rangle \sim \epsilon ^{-1/2}$ for
uniformly-distributed reinjection probabilities below tangency, whereas the scaling breaks down in favor of a logarithmic dependence if such reinjection occurs within a small neighborhood of the tangency region. If the reinjection probability is
delta-distributed, the exponent of the dynamics is found again and rather
generally to be $1/2$ if the distribution is located below and sufficiently
far from tangency, whereas an $\epsilon ^{-1/4}$ scaling is found if the
reinjection is deterministically performed at tangency. Finally, reinjections above
tangency generate trivial $\epsilon ^{0}$ scaling \cite{kim}. On the other hand, natural
reinjection, such as in the case of the logistic map close to any window
of periodicity, occurs due to the presence of homoclinic orbits that take
trajectories that leave the channel to place them arbitrarily close to the
entrance of it at later, unpredictable, times. In this case the reinjection
probability $p_{0}(y)$ is not controlled by hand and the resulting distribution of laminar sizes is asymmetrically U-shaped \cite{hirsch}. However, non controlled reinjection usually yields again the $\epsilon^{-0.5}$ scaling for the mean length of laminar phases.
\bibitem{note2}
It is well known \cite{schuster, hirsch} that
$P(\ell ;\epsilon )=\epsilon p_{0}(y)(1+\tan^{2}[\pi /2-\sqrt{\epsilon }a\ell ])$ \cite{hirsch}, where $p_{0}(y)$ is the
reinjection probability distribution and $a=\frac{1}{2}\frac{d^2 F^{(3)}(x)}{dx^2}|_{x^*,\mu_c}\cdot\frac{d F^{(3)}(x)}{d\mu}|_{x^*,\mu_c}=68.5$ ,
such that $\langle\ell \rangle=\frac{\pi\epsilon^{-0.5}}{2\sqrt a}$ for small values of $\epsilon$.
\end{thebibliography}

\end{document}